\documentclass[trackchanges]{aastex701}

\received{***}
\revised{***}
\accepted{***}

\linenumbers

\begin{document}

\title{Solar Soft X-ray Coronal Dimming in a Failed Eruption Associated with Plasma Cooling}

\author[orcid=0009-0003-5457-4106,gname=Xinyue,sname='Wang']{Xinyue Wang}
\affiliation{School of Earth and Space Sciences, Peking University, Beijing 100871, People's Republic of China; \url{huitian@pku.edu.cn}}
\affiliation{State Key Laboratory of Solar Activity and Space Weather, National Space Science Center, Chinese Academy of Sciences, Beijing 100190, People's Republic of China}
\affiliation{Institute of Physics, University of Graz, A-8010 Graz, Austria}
\email{xinywang@stu.pku.edu.cn}

\author[orcid=0000-0003-2073-002X,gname=Astrid,sname=Veronig]{Astrid M. Veronig}
\affiliation{Institute of Physics, University of Graz, A-8010 Graz, Austria}
\affiliation{Kanzelhöhe Observatory for Solar and Environmental Research, University of Graz, A-9521 Treffen, Austria}
\email{astrid.veronig@uni-graz.at}

\author[orcid=0000-0001-7866-4358,gname=Hechao,sname='Chen']{Hechao Chen}
\affiliation{Department of Astronomy, Key Laboratory of Astroparticle Physics of Yunnan Province, Yunnan University, Kunming 650091, People’s Republic of China}
\email{hechao.chen@ynu.edu.cn}

\author[orcid=0000-0002-1369-1758,gname=Hui, sname='Tian']{Hui Tian} 
\affiliation{School of Earth and Space Sciences, Peking University, Beijing 100871, People's Republic of China; \url{huitian@pku.edu.cn}}
\affiliation{State Key Laboratory of Solar Activity and Space Weather, National Space Science Center, Chinese Academy of Sciences, Beijing 100190, People's Republic of China}
\email{huitian@pku.edu.cn}

\begin{abstract}

Coronal dimmings are observed as sudden and localized reductions in the extreme-ultraviolet and X-ray emission of the solar corona. Traditionally, significant dimmings of spectral lines formed at temperatures of 1--2~MK are regarded as indicators of coronal mass ejections~(CMEs), reflecting the density depletion caused by plasma escaping into interplanetary space. In this Letter, we report a peculiar deep coronal dimming event predominantly observed in high-temperature spectral lines following an M8.8-class confined solar flare associated with a failed filament eruption. Sun-as-a-star measurements from the Geostationary Operational Environmental Satellite and the Extreme Ultraviolet Variability Experiment reveal intensity reductions exceeding 30\% in soft X-ray~(SXR) and measurable decreases in \ion{Fe}{18} (6.5 MK) and \ion{Fe}{20} (9.3 MK). Spatially resolved observations from the Atmospheric Imaging Assembly demonstrate that the dimming originates from the active region core, while the Solar Terrestrial Relations Observatory-A shows no evidence for CME-driven mass loss. Differential emission measure analysis reveals plasma at temperatures $>$5~MK cooling into lower temperatures, supporting plasma cooling as the dominant contributor to the hot-band dimming rather than CME-associated plasma escape. This event demonstrates that deep hot SXR dimmings can occur without substantial CME-driven mass loss, suggesting that alternative physical mechanisms may also account for unresolved stellar dimmings in addition to the commonly inferred CME signatures.

\end{abstract}


\section{Introduction}
Coronal dimming refers to a sudden decrease in coronal emission detected in extreme-ultraviolet~(EUV) and soft X-ray~(SXR) imaging observations. Initially interpreted as transient coronal holes during the Skylab era \citep{1976SoPh...48..381R, 1983SSRv...34...21R}, decades of multi-instrument observations have established a spatial and temporal coincidence between CME-associated dimmings and plasma density depletion \citep{1997ApJ...491L..55S,2000A&A...358.1097H, 2000GeoRL..27.1431T}. It was shown in numerous studies that the decrease of the EUV and SXR intensity observed in coronal dimmings is mostly due to a decrease in density and not due to variations in temperature (e.g., \citealt{1998GeoRL..25.2465T,1999ApJ...520L.139Z,2018ApJ...857...62V}). In addition, distinct correlations have been reported between dimming properties with CME mass and speed (e.g., \citealt{2009ApJ...706..376A, 2019ApJ...874..123D, 2020ApJ...896...17C}). Systematic studies revealed that long-lasting dimmings related to the mass loss of coronal mass ejections (CMEs) are best observed in the 1–2~MK temperature range \citep{2012ApJ...748..106T, 2018ApJ...863..169D, 2018ApJ...857...62V}. This suggests that a substantial fraction of the observed dimming signal arises from plasma depletion in ambient coronal structures surrounding the eruption site, rather than exclusively from the hottest active-region core \citep{2016SoPh..291.1761H}. As they directly trace the global coronal responses, dimming observations have become an indispensable diagnostic for understanding coronal plasma properties and the mechanisms of energy release during large-scale magnetic eruptions \citep{2016ApJ...825...37C, 2016ApJ...830...20M, 2024A&A...691A.195R}. For a more detailed discussion on dimming properties and their relation to CMEs and flares, we refer to the recent review on coronal dimmings by \citet{2025LRSP...22....2V}.

Numerous statistical studies have demonstrated a close relationship between coronal dimmings and CMEs, with the exact numbers depending on the data, event selection and set-up of the different studies \citep{2025LRSP...22....2V}. For example, \citet{2008A&A...478..897B} reported that over 70\% of CMEs in their sample could be traced back to coronal dimming regions identified in either \ion{Mg}{9} or \ion{Fe}{16} lines. Halo CMEs have been reported to be associated with coronal dimmings by at least 50 to 80\% \citep{2008ApJ...674..576R, 2019ApJ...874..123D}. \citet{2021NatAs...5..697V} found that 32 out of 38 CMEs (about 84\%) associated with large flares revealed a dimming in full-Sun EUV fluxes. This solar correspondence has motivated the use of dimmings as a proxy for detecting stellar CMEs \citep{2020IAUS..354..426J,2021NatAs...5..697V, 2022ApJ...936..170L}. However, the stellar CME candidates exhibit an average dimming depth of about 22.7\%, which is nearly an order of magnitude greater than the typical 2\% observed for solar CMEs (\citealt{2021NatAs...5..697V}). This may be a selection effect (since stellar dimmings are more difficult to detect), but it also necessitates further studies of whether such profound dimmings result solely from mass loss or involve additional physical processes.

To date, several other mechanisms such as obscuration by cool prominence materials \citep{2024ApJ...970...60X,2025A&A...695A..12H}, thermal evolution \citep{2010ApJ...720L..88R} or wave-induced density perturbations \citep{2011ApJ...739...89M,2019ApJ...877...68P} have also been identified as causes of dimming. Nevertheless, these effects are often relatively short-lived, may reveal a different spatio-temporal behavior, or may be more prominent in different temperature regimes.

While most studies have focused on eruptive flares accompanied by CMEs, the role of confined flares and failed eruptions in producing coronal dimming remains largely unknown. 
Confined flares are flares that are not associated with a CME (e.g., \citealt{Moore_2001}). Failed eruptions can occur as a subclass of confined flares, where eruptive signatures are observed low in the corona in association with a flare but no CME that escapes to interplanetary space \citep{Ji_2003,Gilbert_2007,Mrozek_2020,2025SCPMA..6879611Z}. They may appear as rapidly evolving magnetic structures that are suddenly stopped in the corona (e.g., \citealt{Ji_2003,Gou_2026}), probably by the overlying magnetic field (e.g., \citealt{Torok_2004,Torok_2005,Thalmann_2015,Li_2020}), highlighting the importance of the three-dimensional coronal magnetic structure for understanding eruption instability and outcome \citep{Jiang_2022}.
\citet{2020A&A...633A.142Z} reported a remote dimming associated with a confined circular-ribbon flare, attributing it to density depletion in large-scale loops. In the sample of 44 large flares (GOES class $\ge$M5) analyzed by \citet{2021NatAs...5..697V}, 6 were confined events. One of them revealed a weak dimming in the SDO/EVE 15-25~nm passband. Crucially, these solar studies primarily targeted low-temperature EUV passbands, whereas stellar dimmings so far are mostly detected at SXRs \citep{2021NatAs...5..697V,Namekata_2024}. For more active stars, the quiescent coronal temperature is higher than that of the solar corona, revealing stronger emission in SXR \citep{2015A&A...578A.129J}. Therefore, it is reasonable to expect dimming to be detected at shorter wavelengths. However, this wavelength discrepancy still underscores the urgent need to extend the search for dimming signatures to higher-temperature channels to bridge the gap between solar and stellar observations.

In this Letter, we report a peculiar coronal dimming predominantly observed in spectral lines formed at high temperatures in a failed eruption. We show that cooling is the most plausible contributor to the high-temperature dimming, while the observations do not support substantial CME-driven mass loss.

\begin{figure*}[ht!]
\plotone{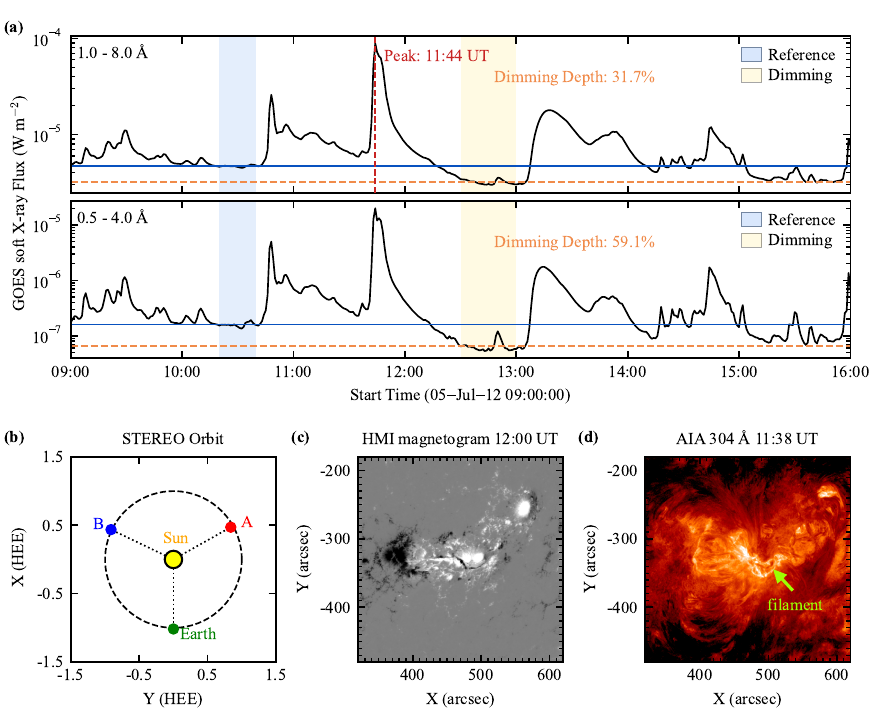}
\caption{Overview of the M8.8 confined flare on 2012 July 05. (a) GOES SXR flux in the 1--8~\AA\ and 0.5--4~\AA\ bands. The blue and yellow shaded regions denote the time intervals used to calculate the reference baseline and dimming depth, respectively. The solid blue line denotes the baseline intensity, while the horizontal red dashed line marks the dimming level. (b) Positions of SDO and STEREO-A in the ecliptic plane. (c) SDO/HMI line-of-sight magnetogram. Magnetic field strength is shown with a color scale saturated at $\pm$1000~G. (d) SDO/AIA 304~\AA\ image of AR 11515, with the filament that erupted in the M8.8 flare indicated by a green arrow.
\label{fig:goes}}
\end{figure*}

\section{Data} \label{sec:Data}

We study the M8.8 flare that occurred on 2012 July 5 in AR 11515 (heliographic position S18°, W32°) with peak time 11:44~UT. The SXR flux was recorded by the Geostationary Operational Environmental Satellite (GOES). As shown in Figure~\ref{fig:goes}(a), the SXR flux exhibits a rapid impulsive phase and subsequently decays below the pre-flare level by approximately 12:15~UT. To analyze the thermal and dynamic evolution of this event, we combined high-resolution stereoscopic imaging observations together with Sun-as-a-star irradiance measurement.

Primary observations were obtained from the Solar Dynamics Observatory (SDO; \citealt{2012SoPh..275....3P}), specifically the Atmospheric Imaging Assembly (AIA; \citealt{2012SoPh..275...17L}) and the EUV Variability Experiment (EVE; \citealt{2012SoPh..275..115W}). AIA provides full-disk EUV images with a spatial resolution of $1\farcs5$ and a 12~s cadence. Its seven coronal channels (131, 94, 335, 211, 193, 171, and 304~\AA) sample plasma from the transition region (0.05~MK) to the hot flaring corona ($>$10~MK). Complementary Sun-as-a-star irradiance data are provided by EVE, which measures the EUV spectrum from 6 to 105~nm with 0.1~nm resolution and 10~s cadence. Specifically, we utilized the EVE Version 8 level~2B lines products, which provide the integrated irradiances extracted from the Multiple EUV Grating Spectrographs spectra at a 60-second cadence.

To assess whether this event is confined and to mitigate projection effects from SDO’s disk-center perspective, we utilized the Extreme Ultraviolet Imager (EUVI; \citealt{2008SSRv..136...67H}) on board the Solar Terrestrial Relations Observatory (STEREO; \citealt{2008SSRv..136....5K}). At the time of the flare, STEREO-A was located approximately 119$^\circ$ west of the Sun–Earth line (Figure~\ref{fig:goes}b), which means that in the STEREO-A view the flare is located close to the limb, approximately at East~87°.
EUVI observes the chromosphere and low corona (up to about 1.7~$R_\odot$) in four EUV passbands centered at 304, 171, 195, and 284~\AA. The spatial scale is $1\farcs6$, and the time cadence for our event is 5~minutes in 195~\AA\ and 10~minutes in 304~\AA.

\section{Results} \label{sec:Results and Analysis}

\subsection{Event Overview\label{subsec:overview}}

The target flare occurs on 2012 July 5, following a smaller M2.5 event in AR 11515. The flare's onset is at 11:39~UT, and the 1--8~\AA\ SXR flux measured by GOES-15 reaches its peak at 11:44~UT. As shown in Figure~\ref{fig:goes}(a), this flare is followed by a uniquely profound dimming in both the GOES~1--8~\AA\ and 0.5--4~\AA\ bands. Using the average flux from 10:20--10:40~UT as the reference baseline, we found that the 1--8~\AA\ and 0.5--4~\AA\ bands exhibited intensity drops of 31.7\% and 59.1\% after the flare, respectively.

\begin{figure*}[ht!]
\plotone{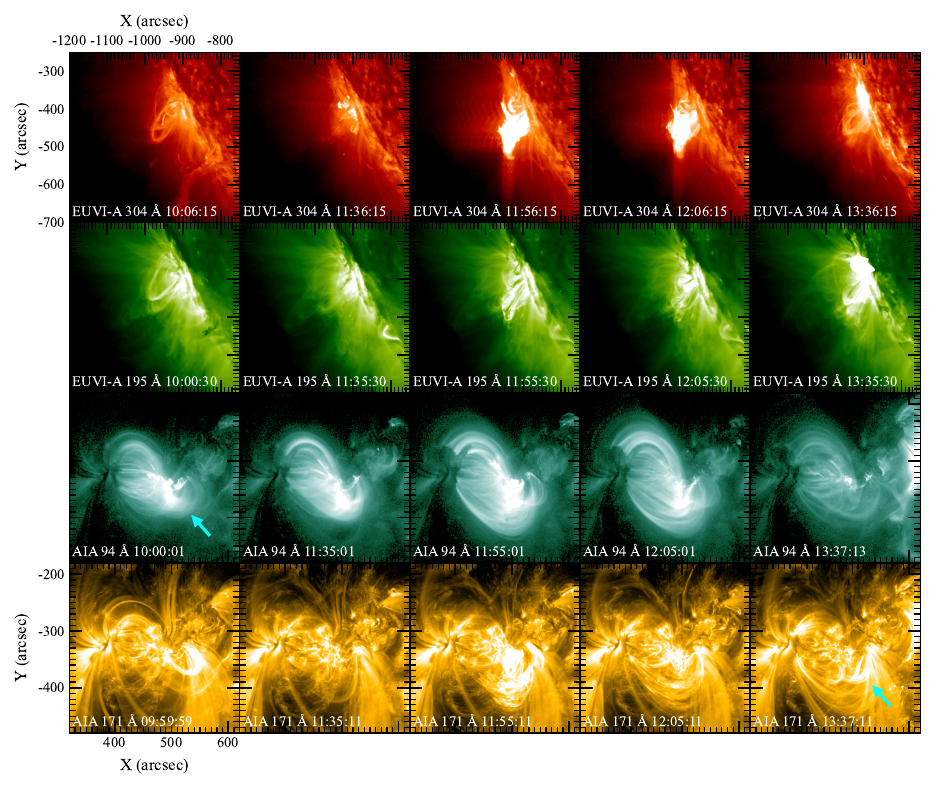}
\caption{Evolution of the M8.8 flare on 2012~July~5. The top and second rows show the side-on view from STEREO-A/EUVI in the 304~\AA\ and 195~\AA\ channels. The third and bottom rows present the disk-center view from SDO/AIA in the 94~\AA\ and 171~\AA\ channels. The blue arrows indicate the pre-flare and cooling active-region loops. An animation covering the time range 10:00~UT to 15:00~UT is available online, which displays the evolution of the eruption and cooling through unannotated AIA and EUVI-A images.
The animation’s duration is 13 s.
\label{fig:eruption}}
\end{figure*}

During the flare, a filament indicated by the arrow in Figure~\ref{fig:goes}(d) rapidly ascends but ceases rising before 12:05~UT. While some of the active-region loops are seen to expand and shift outward in AIA~94~\AA\ during this process, no signatures of escaping plasma are detected above the limb in EUVI-A images~(see Figure~\ref{fig:eruption}). Although STEREO-A coronagraph observations record a material outflow at 11:20~UT, this outflow is attributed to a filament eruption southwest of our active region. No obvious outward-propagating material associated with our event is seen in the STEREO-A coronagraph data. These observations confirm that the event is a confined flare, consistent with its previous classification as a failed eruption by \citet{2019ApJ...881..151L}.

\subsection{Dimming Source Region Identification\label{subsec:Identification}}

\begin{figure*}[ht!]
\plotone{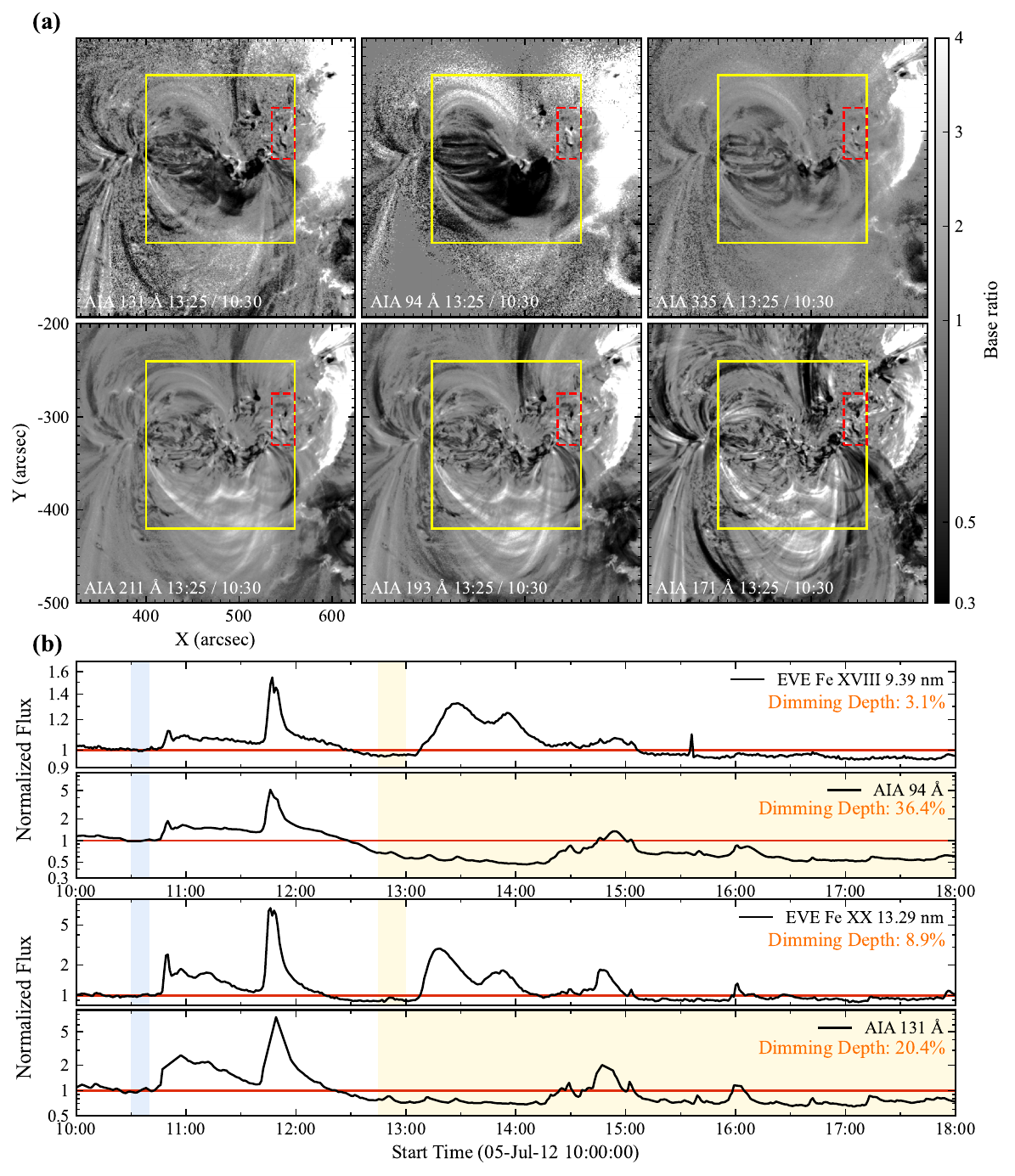}
\caption{Identification of the dimming source region and comparison between AIA and EVE light curves. (a) AIA base-ratio images at 13:25~UT, using 10:30~UT as the reference. The yellow box indicates the region used for extracting the localized AIA light curves. The red dashed box marks a neighboring area excluded from the integration to avoid contamination from an unrelated CME eruption. (b) Comparison of normalized light curves between EVE spectral lines and AIA source-region counts. The light yellow shaded regions denote the time intervals used for calculating the dimming depth. The upper panel compares the EVE \ion{Fe}{18} (9.39~nm) line with the AIA~94~\AA\ channel, while the lower panel compares the EVE \ion{Fe}{20} (13.29~nm) line with the AIA~131~\AA\ channel. An animation showing the evolution of the AIA base-ratio images relative to 10:30~UT is available. The animation covers 10:30--18:00~UT and has a duration of about 19~s.
\label{fig:region}}
\end{figure*}

In order to identify the source region of the dimming signals detected in GOES SXR, we carefully inspected the AIA EUV images and light curves in conjunction with the corresponding EVE spectral intensities. Figure~\ref{fig:region}(a) displays the AIA base-ratio images at 13:25~UT. The AIA light curves were extracted from the region enclosed by the yellow box ($x=[400,560]$ and
$y=[-420,-240]$). The ROI was defined from the pre-event AIA 94~\AA\ morphology to encompass the entire hot-loop system in the eastern part of the active region that participated in the failed eruption. Because the rectangular box also included a small portion of the western part of the active region, the subregion enclosed by the red box ($x=[535,560]$ and $y=[-330,-275]$) was excluded to avoid contamination from a neighboring eruption that occurred there after 13:00~UT. As robustness checks, we repeated the analysis using alternative region selections, including a larger box enclosing the full active region, a shifted box, and a mask selecting pixels above the 85th percentile of the pre-event AIA 94~\AA\ intensity. Although the quantitative dimming depths vary with the exact region selection, all three alternatives reproduce the same qualitative decrease in hot-channel emission. A nearby relatively stable region was additionally examined as a control and shows no comparable decline. In Figure~\ref{fig:region}(a), base-ratio images reveal pronounced dark regions in the 94~\AA\ and 131~\AA\ channels after 12:24~UT (see the online animated version of Figure~\ref{fig:region}), whereas the cooler AIA passbands do not show a comparable prompt dimming signal in the early post-flare phase. This high-temperature selectivity is further evidenced by the EVE spectral lines, where the hot \ion{Fe}{18}~(9.4~nm, 6.5~MK) and \ion{Fe}{20}~(13.3~nm, 9.3~MK) lines exhibit a clear emission reduction after 12:30~UT. As shown in Figure~\ref{fig:region}(b), the temporal evolution of the EVE spectral line irradiance closely tracks the profile of the localized AIA light curve before, during, and after the flare. This temporal and thermal consistency between the Sun-as-a-star EVE observations and the spatially resolved AIA source-region emission indicates that this active-region core was the dominant source of the full-disk dimming signal and that the dimming was likely driven by the evolution of plasma at temperatures above 5~MK.

To quantify this dimming feature, we calculated the dimming depths for both bands and both instruments. The light curves were all normalized to the average flux of the pre-event period, calculated between 10:30--10:40~UT. We used the time range 12:45--18:00~UT to calculate the localized AIA dimming depths, as the dimming remains directly visible in AIA for at least 5 hours. For EVE, we set a rather short window between 12:45--13:00~UT to calculate the dimming depth, since the irradiances are full-disk integrated and we could not differentiate the effects from flares and other variations that occurred outside our region of interest. This yielded dimming depths of 3.1\% for \ion{Fe}{18} and 8.9\% for \ion{Fe}{20} in EVE high-temperature lines, and 36.4\% for 94~\AA\ and 20.4\% for 131~\AA\ in the localized AIA source-region light curves. The dependence of the inferred dimming depth on the choice of reference interval is discussed further in Appendix~\ref{sec:appendixA}. The close similarity between these high-temperature light curves and the GOES SXR flux profiles indicates that the SXR dimming originates from the flare site.

\subsection{Thermal Evolution\label{subsec:thermal}}

Unlike eruptive flares, confined flares are not expected to produce substantial mass loss into the heliosphere. As shown in Figure~\ref{fig:eruption}, the morphology and spatial location of the loop system suggest a dramatic thermal evolution following the flare. At 10:00~UT, the active-region loops are visible in the AIA~94~\AA\ channel, while no corresponding structures are identifiable in the cooler 171~\AA\ channel. During the flare, a subset of these loops pushed outward by the rising filament are clearly seen at 11:55~UT. Approximately two hours after the flare peak, this overlying structure substantially fades in the 94~\AA\ passband, but becomes prominently visible in AIA~171~\AA. This evolution indicates a substantial plasma cooling from over 5~MK to approximately 1~MK within these closed magnetic structures.

\begin{figure*}[ht!]
\plotone{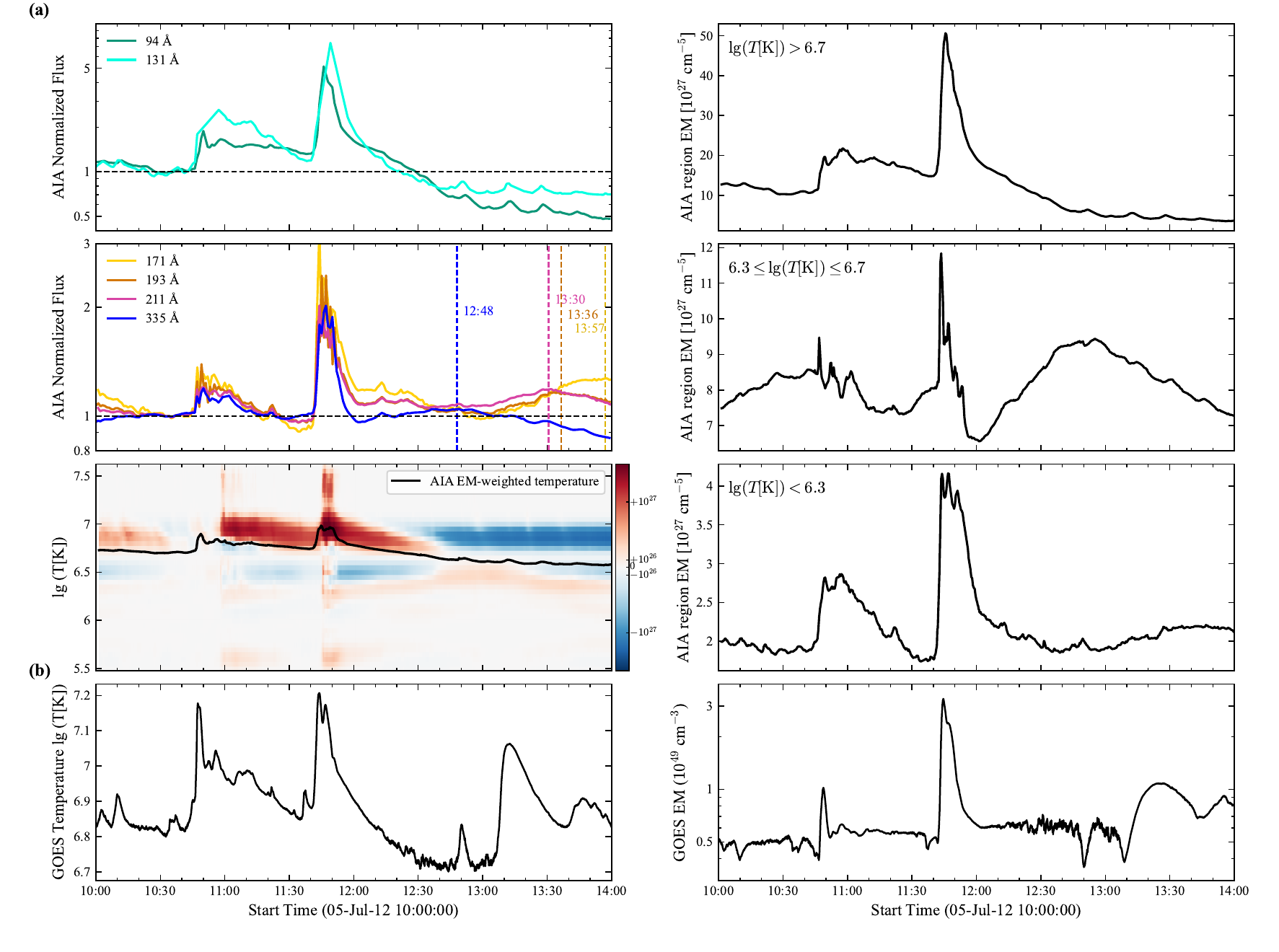}
\caption{Thermal evolution and plasma cooling during the confined flare. All AIA quantities are derived from the same region as in Figure~\ref{fig:region}. In panel (a), the left column shows the normalized AIA light curves in the hot channels (94 and 131~\AA) and cooler channels (171, 193, 211, and 335~\AA), followed by the evolution of the base-difference EM distribution. Red and blue indicate increases and decreases in EM, respectively, relative to the pre-flare mean during 10:30--10:40~UT. The right column shows the regional AIA EM integrated over the high-, intermediate-, and low-temperature ranges. Panel (b) presents the temperature and EM derived from the GOES soft X-ray fluxes.
\label{fig:lightcurve}}
\end{figure*}

To further investigate the thermal evolution of the event, we performed a differential emission measure (DEM) analysis for the flare region based on the methods of \citet{2018ApJ...856L..17S} and \citet{2015ApJ...807..143C}. The available Hinode/XRT observations did not provide sufficiently co-temporal multi-filter coverage with a stable field of view. Therefore, the inversion used only the six optically thin coronal AIA channels. The DEM inversion was performed independently for each pixel. The resulting EM distributions were first spatially averaged over the selected ROI, and the reference distribution was then obtained by averaging these ROI-averaged distributions over 10:30–10:40 UT. The base-difference EM distribution was obtained by subtracting this reference distribution from the original distribution at each time. Figure~\ref{fig:lightcurve}(a) shows the time evolution of the base-difference EM distribution and the EM-weighted temperature derived from the original EM distribution. During the M8.8-class flare, the EM increases across all temperatures, with a pronounced enhancement~($\sim 4\times 10^{28}\mathrm{cm^{-5}}$) at high temperatures~($>$5~MK). After the flare peak, the hot plasma cools progressively, manifesting as a positive base-difference EM feature (red) that migrates from high to low temperatures over time. To further quantify this evolution, we integrated the regional AIA EM over three temperature ranges, as shown in the right column of Figure~\ref{fig:lightcurve}(a). After the flare peak, the high-temperature EM ($\log T[\mathrm{K}]>6.7$) decreases substantially, whereas the intermediate-temperature EM ($6.3\leq\log T[\mathrm{K}]\leq6.7$) subsequently increases. The low-temperature EM ($\log T[\mathrm{K}]<6.3$) also shows a later enhancement. Given that the DEM inversion is constrained by only six AIA coronal channels and uses a relatively fine temperature grid, the fine-scale bin-to-bin variations in the base-difference EM should not be interpreted as fully resolved physical temperature structures. We therefore base our interpretation primarily on the EM evolution integrated over broad temperature ranges. The temporal redistribution of EM from high to lower temperatures further supports progressive plasma cooling within the confined loop system.

We also derived the AIA 335, 211, 193 and 171~\AA\ light curves averaged over the same region where we extracted the localized 94 and 131~\AA\ light curves. The negative temperature drift pattern from high to low temperatures is consistent with the sequential peaks observed in the AIA light curves (see Figure~\ref{fig:lightcurve}). After the flare peak, the AIA~335~\AA\ ($\sim$2.5~MK) flux reaches its maximum first at 12:48~UT, followed sequentially by 211~\AA\ ($\sim$2.0~MK, 13:30~UT), 193~\AA\ ($\sim$1.6~MK, 13:36~UT), and finally 171~\AA\ ($\sim$0.6~MK, 13:57~UT). The simultaneous darkening of high-temperature channels (94 and 131~\AA) and the brightening of lower-temperature channels provide strong evidence for plasma cooling. The EM-weighted temperature decreases from about 5 MK during the reference phase to about 4 MK after the M8.8 flare. As an independent full-disk diagnostic, the temperature derived from the two GOES SXR channels decreases from approximately 6.5 MK at around 10:30 UT to approximately 5.2 MK at around 12:30 UT. A fixed pre-flare background subtraction is not applicable in this case because the observed GOES flux falls below the pre-flare level during the subsequent dimming phase, leading to negative background-subtracted fluxes and invalid temperature and emission measure estimates. The GOES-derived temperature and EM are treated only as full-disk isothermal-equivalent supporting diagnostics. These results also suggest that plasma cooling, rather than large-scale plasma evacuation, is the dominant contributor to the dimming in this confined event.

\subsection{Pre-flare Thermal State}\label{subsec:pre-flare}
NOAA active region 11515 is complex and highly active, producing nearly 100 flares and at least 20 white-light flares during its passage across the solar disk (\citealt{2018A&A...613A..69S}). As shown in Figure~\ref{fig:eruption}, this active region exhibits strong emission in AIA~94~\AA\ on July~5. To determine whether this emission represented a transient hot state immediately preceding the dimming or an intrinsic property of the active region itself, we examined AIA images one day before the flare.

\begin{figure}[ht!]
\plotone{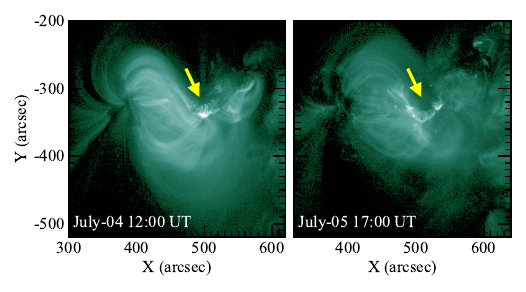}
\caption{AIA~94~\AA\ images showing persistent high-temperature loop emission in the core of NOAA 11515. The left panel (2012~July~4, 12:00 UT) displays the active region approximately one day prior to the M8.8 flare. The right panel (2012~July~5, 17:00 UT) shows the post-flare stage. Yellow arrows indicate active-region core loops where recurrent minor flares were observed.
\label{fig:heating}}
\end{figure}

Figure~\ref{fig:heating} shows that the eastern part of AR 11515 exhibits persistent high-temperature emission by 12:00~UT on July~4. The active-region core loops indicated by the yellow arrow remain visible from July~4 to July~5 even after the flare. At the western footpoint of these loops, recurrent minor flares were observed. Situated beneath this flare site, a narrow ribbon of negative-polarity photospheric flux is embedded within the dominant positive-polarity flux of the leading sunspot in AR 11515 (see Figure~\ref{fig:goes}(c)). This magnetic setting—a parasitic polarity enclosed by a unipolar field—is a highly favorable environment for the initiation of recurrent eruptive events in AR 11515 (\citealt{2021ApJ...911...33C}). This suggests ongoing heating capable of maintaining the plasma at temperatures of about 5~MK, implying that the strong hot emission prior to the flare is not a transient feature.

\section{Summary and Discussion} \label{sec:Discussion}

In this Letter, we report a pronounced SXR coronal dimming following an M8.8 confined flare and failed filament eruption. By combining Sun-as-a-star GOES and EVE irradiance measurements with spatially resolved AIA observations, we demonstrate that the dimming originated directly from the active region core and occurred in the absence of CME-driven mass loss. The SXR flux decreases by over 30\%, while EVE Fe XVIII and Fe XX lines show measurable irradiance reductions by 3\% and 9\%, respectively, consistent with the strong localized dimming observed in AIA~94~\AA\ and 131~\AA.

Previous solar studies have established a close relationship between coronal dimmings and CMEs. In eruptive events, dimmings are typically most prominent in spectral lines with formation temperatures of 1--2~MK and often appear multi-thermally across cooler EUV passbands, consistent with density depletion caused either by CME-associated mass loss or by the expansion and stretching of overlying magnetic fields (e.g., \citealt{2011ApJ...739...59W,2016ApJ...830...20M,2018ApJ...857...62V,2018ApJ...863..169D,2024ScChE..67.1592L}). The spatio-temporal relation between dimming regions and flare ribbons can further provide information on the magnetic flux systems involved in the eruption process, as discussed by \citet{2025LRSP...22....2V}. By contrast, observations and studies of dimming in confined flares are scarce. The few reported cases included a remote dimming in a confined circular-ribbon flare reported by \citet{2020A&A...633A.142Z}, and a weak dimming in the full-disk EVE light curve for one of the confined flares studied by \citet{2021NatAs...5..697V}. The dimming reported in this Letter is different in two respects: it was a pronounced hot-band dimming occurring in a confined flare without an associated CME, and it originated from the active-region core rather than from a remote region. This core origin is demonstrated by the region-selected AIA light curves, whose intensity decrease closely matches the EVE \ion{Fe}{18} and \ion{Fe}{20} irradiance decreases, indicating that the Sun-as-a-star hot-line dimming is produced by the confined flare source region itself. Unlike the standard CME-associated dimming pattern, the dimming in our event is strongly temperature-selective, with decreases observed in AIA 94~\AA\ and 131~\AA\ and in EVE \ion{Fe}{18} and \ion{Fe}{20} lines, while remaining negligible in the cooler AIA channels in the early post-flare phase. This argues against the canonical 1--2 MK mass-depletion picture typical of CME-associated dimmings. The dimming depth measured by GOES is even larger. This may partly reflect differences between the evolution of broadband SXR emission, including its continuum component, and that of individual EUV lines during plasma cooling, as well as differences in the contribution of the target active region to the respective full-disk signals.

Thermodynamic mechanisms have previously been invoked to explain dimming-like signatures without invoking mass loss. As proposed by \citet{2010ApJ...720L..88R} and summarized by \citet{2014ApJ...789...61M}, this process is typically characterized by emission decrease in lines with lower peak formation temperatures and a near-simultaneous emission increase in lines with higher temperatures during heating phases; vice versa for cooling. In our analysis, the emission in cooler passbands exhibited a clear cooling sequence following the flare, with AIA~335~\AA\ peaking at~12:48, 211~\AA\ at~13:30, 193~\AA\ at~13:36 and 171~\AA\ at~13:57, respectively. Notably, after the main flare peak, the AIA 94 \AA\ emission exhibited a more gradual decay than the hotter diagnostics (e.g., AIA~131~\AA\ and GOES SXRs). This suggests that the initially hotter plasma cooled through the temperature response range of the AIA~94~\AA\ channel.

As shown in Figure \ref{fig:heating}, the active region exhibited persistent high-temperature emission in AIA~94~\AA\ at least one day before the M8.8 flare. This suggests that the active region is capable of heating and maintaining active-region loops at high temperature ($>$5 MK). Although the hot-loop system exists well before the event, its displacement during the failed eruption, followed by its fading in the hot AIA channels and ordered appearance in progressively cooler channels, temporally and morphologically links the observed dimming to the M8.8 flare rather than to gradual active-region evolution. A plausible interpretation for this unusual coronal dimming in hot bands may involve an intense chromospheric evaporation phase, which can rapidly inject hot and dense plasma into coronal loops (e.g., \citealt{2015ApJ...811..139T,S22593-lijianping-F}). Similar flare-driven hot coronal flows and density enhancements have also been reported in X-ray spectroscopy of an active M dwarf \citep{2022ApJ...933...92C}. Such a density enhancement could substantially increase radiative losses, which are approximately proportional to the square of the electron density \citep{1966ApJ...144..244T,1970A&A.....6..468L,1971ApJ...168..283T}. Once the heating rate becomes insufficient to balance the enhanced radiative losses, the loops enter a cooling phase, manifesting as the profound dimming observed in high-temperature passbands.

The DEM analysis broadly supports this interpretation, although it does not provide a closed thermal or mass budget. The ROI-averaged coronal EM constrained by the six AIA channels decreases from approximately $2.0\times10^{28}\ \mathrm{cm}^{-5}$ during 10:30--10:40~UT to approximately $1.3\times10^{28}\ \mathrm{cm}^{-5}$ during 13:50--14:00~UT. This corresponds to a reduction of about $7\times10^{27}\ \mathrm{cm}^{-5}$, or approximately 35\%. However, coronal-rain-like condensations become visible in the AIA 304~\AA\ channel after the confined eruption, as shown in the animation associated with Figure~\ref{fig:eruption}. Part of the emitting plasma may therefore have cooled below the temperature range reliably constrained by the six coronal AIA channels. In addition, the cool condensations may introduce absorption and optical-depth effects that are not accounted for by the optically thin coronal DEM inversion. Consequently, the measured decrease in total coronal EM cannot be interpreted straightforwardly as an equivalent loss of plasma from the closed loop system. Instead, the combined evolution of the temperature-resolved EM, the migration of emission from hotter to cooler temperature ranges, and the subsequent appearance of cool condensations suggest that plasma cooling is the dominant, or at least the most plausible, contributor to the observed hot-channel dimming, although some reduction of the EM within the AIA-sensitive coronal temperature range also occurred.

Numerical simulations of magnetically confined eruptions in M-dwarf coronae by \citet{2019ApJ...884L..13A} have demonstrated that confined flares could also produce dimmings. Crucially, they proposed that such dimming signatures could be driven by thermodynamic changes. However, we note that in their case the flare signature is produced by compression by the suppressed (failed) eruption interacting with the overlying structures causing the heating, whereas in our case the flare energy release is due to magnetic reconnection and not due to compression at the failed eruption fronts.

Sun-as-a-star solar observations have demonstrated that full-disk EUV irradiance can capture eruptive signatures and measurable coronal dimmings associated with CME-related mass loss \citep{2011ApJ...739...59W,2016ApJ...830...20M,2022ApJ...931...76X}, thereby providing an important bridge between spatially resolved solar events and unresolved stellar observations \citep{2021NatAs...5..697V,2022SerAJ.205....1L,2025kiss.rept.....L}. Our results add a solar perspective to this framework by showing that a confined flare can also produce a deep and long-lasting dimming in hot channels, especially observed in SXRs. This finding does not diminish the usefulness of dimming as a CME diagnostic but suggests that, in unresolved stellar observations, the coronal dimming observed at high temperatures may reflect a combination of plasma evacuation and thermal evolution. Accordingly, interpretations of stellar hot-band dimmings may benefit from taking temperature effects into account, ideally together with complementary diagnostics. Because this study examines only a single event, the frequency and properties of such hot-band dimmings in confined flares remain to be established through future statistical investigations.

\appendix 
\setcounter{figure}{0}
\renewcommand{\thefigure}{A\arabic{figure}}

\section{Long-term evolution and dependence on the reference level}\label{sec:appendixA}

\begin{figure}[ht!]
\plotone{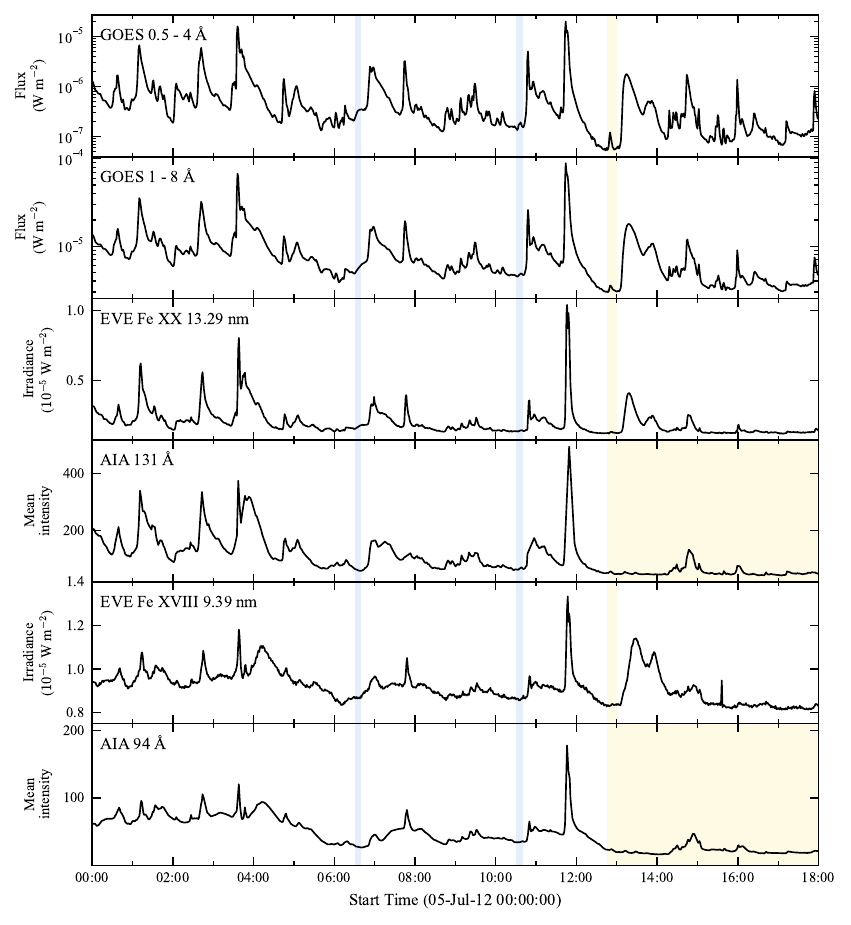}
\caption{Long-term light curves of GOES 0.5–4~\AA, GOES 1–8~\AA, EVE \ion{Fe}{20} 13.29 nm, AIA 131~\AA, EVE \ion{Fe}{18} 9.39 nm, and AIA 94~\AA, shown from top to bottom. The blue shaded intervals indicate the alternative reference intervals. The yellow shaded interval denotes 12:45–13:00 UT for the GOES and EVE measurements, while the corresponding AIA panels show the extended local-dimming intervals.
\label{fig:appendix1}}
\end{figure}

Because NOAA AR 11515 is highly active during this period, we show the light curves over a longer time interval in Figure~\ref{fig:appendix1} and examine how the inferred dimming depths depend on the adopted pre-event reference level. The AIA 94 and 131~\AA\ light curves were calculated by averaging all valid pixels within the fixed region shown in Fig.~\ref{fig:region}(a), after excluding the small subregion affected by unrelated emission. Since our primary interest is the dimming region itself, the reference intervals were selected mainly according to periods during which the local AIA emission showed no prominent eruptive enhancement. By contrast, the GOES and EVE measurements represent full-disk emission and may be affected by activity elsewhere on the Sun. Their reference levels and derived dimming depths should therefore be regarded as supporting estimates rather than measurements exclusively associated with NOAA AR 11515.

We considered three possible reference levels: (1) the mean intensity or flux over 06:30--06:40~UT, (2) the mean over 10:30--10:40~UT, and (3) the arithmetic mean of these two independently determined reference levels. We note that the GOES dimming depths quoted in Section~\ref{subsec:overview} and Figure~\ref{fig:goes}(a) were calculated using a slightly broader reference interval of 10:20–10:40 UT. This interval was adopted to reduce the sensitivity of the GOES reference level to a small short-term fluctuation around 10:30 UT. For consistency in the subsequent multi-instrument comparison, we adopted 10:30--10:40~UT as the common reference interval. For each reference choice, the dimming depth was calculated as
\begin{equation}
D = \frac{I_{\mathrm{ref}}-I_{\mathrm{dim}}}{I_{\mathrm{ref}}}\times 100
\end{equation}
where $I_{\mathrm{dim}}$ is the mean intensity or flux over 12:45--13:00~UT. When the mean of the two reference levels is adopted, the resulting dimming depths are 70.4\% and 37.4\% in the GOES 0.5--4 and 1--8~\AA\ bands, 16.8\% and 3.5\% in the EVE \ion{Fe}{20} and \ion{Fe}{18}, and 19.5\% and 27.5\% in the AIA 131 and 94~\AA\ channels, respectively. Using the two reference intervals separately gives corresponding ranges of 57.0--77.4\%, 32.0--42.0\%, 8.9--23.5\%, 3.1--3.9\%, 16.7--22.1\%, and 18.5--34.7\%.

For the spatially resolved AIA measurements, we additionally calculated the mean dimming depth over the extended interval from 12:45 to 18:00~UT. When the mean of the two reference levels is adopted, the corresponding dimming depths are 17.7\% for AIA 131~\AA\ and 29.4\% for AIA 94~\AA. The two individual reference intervals yield ranges of 14.9--20.4\% and 20.7--36.4\%, respectively. The quoted dimming depths should be regarded as reference-dependent estimates rather than measurements relative to a uniquely defined quiet state. Despite the quantitative sensitivity of the inferred dimming depths to the adopted baseline, all three reference choices consistently indicate a reduction in the hot-coronal emission from the selected active-region loops after the eruption.

\begin{acknowledgments}
This work is supported by the National Natural Science Foundation of China (Grant No.~12425301 \& No.~12573061) and the Specialized Research Fund for State Key Laboratory of Solar Activity and Space Weather. We would like to acknowledge the data use from GOES, SDO, and STEREO/SECCHI. SDO is a mission of NASA’s Living With a Star Program. Data from STEREO/SECCHI are produced by the consortium of RAL (UK), NRL (USA), LMSAL (USA), GSFC (USA), MPS (Germany), CSL (Belgium), IOTA (France), and IAS (France). We thank for the technical support of the National Large Scientific and Technological Infrastructure “Earth System Numerical Simulation Facility”. H.T. is also supported by the New Cornerstone Science Foundation through the Xplorer Prize. X.W. also acknowledges the support from the China Scholarship Council (No. 202406010246). H. C. is also supported by the Young Talent Special of the Xingdian Talent Support Program (Grant No. XDYC-QNRC-2023-0255).
\end{acknowledgments}

\bibliography{main}

@ARTICLE{1983SSRv...34...21R,
       author = {{Rust}, D.~M.},
        title = "{Coronal Disturbances and Their Terrestrial Effects}",
      journal = {\ssr},
         year = 1983,
        month = jan,
       volume = {34},
       number = {1},
        pages = {21-36},
          doi = {10.1007/BF00221193},
       adsurl = {https://ui.adsabs.harvard.edu/abs/1983SSRv...34...21R}
}

@ARTICLE{2009ApJ...706..376A,
       author = {{Aschwanden}, Markus J. and {Nitta}, Nariaki V. and {Wuelser}, Jean-Pierre and {Lemen}, James R. and {Sandman}, Anne and {Vourlidas}, Angelos and {Colaninno}, Robin C.},
        title = "{First Measurements of the Mass of Coronal Mass Ejections from the EUV Dimming Observed with STEREO EUVI A+B Spacecraft}",
      journal = {\apj},
         year = 2009,
        month = nov,
       volume = {706},
       number = {1},
        pages = {376-392},
          doi = {10.1088/0004-637X/706/1/376},
       adsurl = {https://ui.adsabs.harvard.edu/abs/2009ApJ...706..376A}
}

@ARTICLE{2019ApJ...874..123D,
       author = {{Dissauer}, K. and {Veronig}, A.~M. and {Temmer}, M. and {Podladchikova}, T.},
        title = "{Statistics of Coronal Dimmings Associated with Coronal Mass Ejections. II. Relationship between Coronal Dimmings and Their Associated CMEs}",
      journal = {\apj},
         year = 2019,
        month = apr,
       volume = {874},
       number = {2},
          eid = {123},
        pages = {123},
          doi = {10.3847/1538-4357/ab0962},
archivePrefix = {arXiv},
       eprint = {1810.01589},
 primaryClass = {astro-ph.SR},
       adsurl = {https://ui.adsabs.harvard.edu/abs/2019ApJ...874..123D}
}

@ARTICLE{2020ApJ...896...17C,
       author = {{Chikunova}, Galina and {Dissauer}, Karin and {Podladchikova}, Tatiana and {Veronig}, Astrid M.},
        title = "{Coronal Dimmings Associated with Coronal Mass Ejections on the Solar Limb}",
      journal = {\apj},
         year = 2020,
        month = jun,
       volume = {896},
       number = {1},
          eid = {17},
        pages = {17},
          doi = {10.3847/1538-4357/ab9105},
archivePrefix = {arXiv},
       eprint = {2005.03348},
 primaryClass = {astro-ph.SR},
       adsurl = {https://ui.adsabs.harvard.edu/abs/2020ApJ...896...17C}
}

@ARTICLE{2008ApJ...674..576R,
       author = {{Reinard}, A.~A. and {Biesecker}, D.~A.},
        title = "{Coronal Mass Ejection-Associated Coronal Dimmings}",
      journal = {\apj},
         year = 2008,
        month = feb,
       volume = {674},
       number = {1},
        pages = {576-585},
          doi = {10.1086/525269},
       adsurl = {https://ui.adsabs.harvard.edu/abs/2008ApJ...674..576R}
}

@ARTICLE{2012ApJ...748..106T,
       author = {{Tian}, Hui and {McIntosh}, Scott W. and {Xia}, Lidong and {He}, Jiansen and {Wang}, Xin},
        title = "{What can We Learn about Solar Coronal Mass Ejections, Coronal Dimmings, and Extreme-ultraviolet Jets through Spectroscopic Observations?}",
      journal = {\apj},
         year = 2012,
        month = apr,
       volume = {748},
       number = {2},
          eid = {106},
        pages = {106},
          doi = {10.1088/0004-637X/748/2/106},
archivePrefix = {arXiv},
       eprint = {1201.2204},
 primaryClass = {astro-ph.SR},
       adsurl = {https://ui.adsabs.harvard.edu/abs/2012ApJ...748..106T}
}

@ARTICLE{1998GeoRL..25.2465T,
       author = {{Thompson}, B.~J. and {Plunkett}, S.~P. and {Gurman}, J.~B. and {Newmark}, J.~S. and {St. Cyr}, O.~C. and {Michels}, D.~J.},
        title = "{SOHO/EIT observations of an Earth-directed coronal mass ejection on May 12, 1997}",
      journal = {\grl},
         year = 1998,
        month = jul,
       volume = {25},
       number = {14},
        pages = {2465-2468},
          doi = {10.1029/98GL50429},
       adsurl = {https://ui.adsabs.harvard.edu/abs/1998GeoRL..25.2465T}
}

@ARTICLE{1999ApJ...520L.139Z,
       author = {{Zarro}, Dominic M. and {Sterling}, Alphonse C. and {Thompson}, Barbara J. and {Hudson}, Hugh S. and {Nitta}, Nariaki},
        title = "{SOHO EIT Observations of Extreme-Ultraviolet ``Dimming'' Associated with a Halo Coronal Mass Ejection}",
      journal = {\apjl},
         year = 1999,
        month = aug,
       volume = {520},
       number = {2},
        pages = {L139-L142},
          doi = {10.1086/312150},
       adsurl = {https://ui.adsabs.harvard.edu/abs/1999ApJ...520L.139Z}
}

@ARTICLE{1976SoPh...48..381R,
       author = {{Rust}, D.~M. and {Hildner}, E.},
        title = "{Expansion of an X-ray coronal arch into the outer corona.}",
      journal = {\solphys},
         year = 1976,
        month = jun,
       volume = {48},
       number = {2},
        pages = {381-387},
          doi = {10.1007/BF00152003},
       adsurl = {https://ui.adsabs.harvard.edu/abs/1976SoPh...48..381R}
}

@ARTICLE{2000A&A...358.1097H,
       author = {{Harrison}, R.~A. and {Lyons}, M.},
        title = "{A spectroscopic study of coronal dimming associated with a coronal mass ejection}",
      journal = {\aap},
         year = 2000,
        month = jun,
       volume = {358},
        pages = {1097-1108},
       adsurl = {https://ui.adsabs.harvard.edu/abs/2000A&A...358.1097H}
}

@ARTICLE{2025LRSP...22....2V,
       author = {{Veronig}, Astrid M. and {Dissauer}, Karin and {Kliem}, Bernhard and {Downs}, Cooper and {Hudson}, Hugh S. and {Jin}, Meng and {Osten}, Rachel and {Podladchikova}, Tatiana and {Prasad}, Avijeet and {Qiu}, Jiong and {Thompson}, Barbara and {Tian}, Hui and {Vourlidas}, Angelos},
        title = "{Coronal dimmings and what they tell us about solar and stellar coronal mass ejections}",
      journal = {Living Reviews in Solar Physics},
         year = 2025,
        month = jul,
       volume = {22},
       number = {1},
          eid = {2},
        pages = {2},
          doi = {10.1007/s41116-025-00041-4},
archivePrefix = {arXiv},
       eprint = {2505.19228},
 primaryClass = {astro-ph.SR},
       adsurl = {https://ui.adsabs.harvard.edu/abs/2025LRSP...22....2V}
}

@ARTICLE{2016ApJ...825...37C,
       author = {{Cheng}, J.~X. and {Qiu}, J.},
        title = "{The Nature of CME-flare-Associated Coronal Dimming}",
      journal = {\apj},
         year = 2016,
        month = jul,
       volume = {825},
       number = {1},
          eid = {37},
        pages = {37},
          doi = {10.3847/0004-637X/825/1/37},
archivePrefix = {arXiv},
       eprint = {1604.05443},
 primaryClass = {astro-ph.SR},
       adsurl = {https://ui.adsabs.harvard.edu/abs/2016ApJ...825...37C}
}

@INPROCEEDINGS{2020IAUS..354..426J,
       author = {{Jin}, M. and {Cheung}, M.~C.~M. and {DeRosa}, M.~L. and {Nitta}, N.~V. and {Schrijver}, C.~J. and {France}, K. and {Kowalski}, A. and {Mason}, J.~P. and {Osten}, R.},
        title = "{Coronal dimming as a proxy for stellar coronal mass ejections}",
    booktitle = {Solar and Stellar Magnetic Fields: Origins and Manifestations},
         year = 2020,
       editor = {{Kosovichev}, A. and {Strassmeier}, S. and {Jardine}, M.},
       series = {IAU Symposium},
       volume = {354},
        month = jan,
        pages = {426-432},
          doi = {10.1017/S1743921320000575},
archivePrefix = {arXiv},
       eprint = {2002.06249},
 primaryClass = {astro-ph.SR},
       adsurl = {https://ui.adsabs.harvard.edu/abs/2020IAUS..354..426J}
}

@ARTICLE{2015A&A...578A.129J,
       author = {{Johnstone}, C.~P. and {G{\"u}del}, M.},
        title = "{The coronal temperatures of low-mass main-sequence stars}",
      journal = {\aap},
         year = 2015,
        month = jun,
       volume = {578},
          eid = {A129},
        pages = {A129},
          doi = {10.1051/0004-6361/201425283},
archivePrefix = {arXiv},
       eprint = {1505.00643},
 primaryClass = {astro-ph.SR},
       adsurl = {https://ui.adsabs.harvard.edu/abs/2015A&A...578A.129J}
}

@ARTICLE{2018A&A...613A..69S,
       author = {{Song}, Y.~L. and {Tian}, H. and {Zhang}, M. and {Ding}, M.~D.},
        title = "{Observations of white-light flares in NOAA active region 11515: high occurrence rate and relationship with magnetic transients}",
      journal = {\aap},
         year = 2018,
        month = jun,
       volume = {613},
          eid = {A69},
        pages = {A69},
          doi = {10.1051/0004-6361/201731817},
archivePrefix = {arXiv},
       eprint = {1801.04371},
 primaryClass = {astro-ph.SR},
       adsurl = {https://ui.adsabs.harvard.edu/abs/2018A&A...613A..69S}
}

@ARTICLE{2019ApJ...877...68P,
       author = {{Podladchikova}, Tatiana and {Veronig}, Astrid M. and {Dissauer}, Karin and {Temmer}, Manuela and {Podladchikova}, Olena},
        title = "{Three-dimensional Reconstructions of Extreme-ultraviolet Wave Front Heights and Their Influence on Wave Kinematics}",
      journal = {\apj},
         year = 2019,
        month = jun,
       volume = {877},
       number = {2},
          eid = {68},
        pages = {68},
          doi = {10.3847/1538-4357/ab1b3a},
archivePrefix = {arXiv},
       eprint = {1904.09427},
 primaryClass = {astro-ph.SR},
       adsurl = {https://ui.adsabs.harvard.edu/abs/2019ApJ...877...68P}
}

@ARTICLE{2025A&A...695A..12H,
       author = {{Hou}, Zhenyong and {Tian}, Hui and {Yan}, Jingye and {Madjarska}, Maria S. and {Zhang}, Jiale and {Xu}, Yu and {Chen}, Hechao and {Wu}, Zhao and {Wu}, Lin and {Lv}, Xuning and {Yang}, Yang and {Liu}, Yujie and {Deng}, Li and {Feng}, Li and {Qiu}, Ye},
        title = "{Radio dimming associated with filament eruptions in the meter and decimeter wavebands}",
      journal = {\aap},
         year = 2025,
        month = mar,
       volume = {695},
          eid = {A12},
        pages = {A12},
          doi = {10.1051/0004-6361/202453282},
archivePrefix = {arXiv},
       eprint = {2504.08391},
 primaryClass = {astro-ph.SR},
       adsurl = {https://ui.adsabs.harvard.edu/abs/2025A&A...695A..12H}
}

@ARTICLE{2022ApJ...936..170L,
       author = {{Loyd}, R.~O. Parke and {Mason}, James Paul and {Jin}, Meng and {Shkolnik}, Evgenya L. and {France}, Kevin and {Youngblood}, Allison and {Villadsen}, Jackie and {Schneider}, Christian and {Schneider}, Adam C. and {Llama}, Joe and {Ramiaramanantsoa}, Tahina and {Richey-Yowell}, Tyler},
        title = "{Constraining the Physical Properties of Stellar Coronal Mass Ejections with Coronal Dimming: Application to Far-ultraviolet Data of ϵ Eridani}",
      journal = {\apj},
         year = 2022,
        month = sep,
       volume = {936},
       number = {2},
          eid = {170},
        pages = {170},
          doi = {10.3847/1538-4357/ac80c1},
archivePrefix = {arXiv},
       eprint = {2207.05115},
 primaryClass = {astro-ph.SR},
       adsurl = {https://ui.adsabs.harvard.edu/abs/2022ApJ...936..170L}
}

@ARTICLE{2000GeoRL..27.1431T,
       author = {{Thompson}, B.~J. and {Cliver}, E.~W. and {Nitta}, N. and {Delann{\'e}e}, C. and {Delaboudini{\`e}re}, J.-P.},
        title = "{Coronal dimmings and energetic CMEs in April-May 1998}",
      journal = {\grl},
         year = 2000,
        month = may,
       volume = {27},
       number = {10},
        pages = {1431-1434},
          doi = {10.1029/1999GL003668},
       adsurl = {https://ui.adsabs.harvard.edu/abs/2000GeoRL..27.1431T}
}

@ARTICLE{2016ApJ...830...20M,
       author = {{Mason}, James Paul and {Woods}, Thomas N. and {Webb}, David F. and {Thompson}, Barbara J. and {Colaninno}, Robin C. and {Vourlidas}, Angelos},
        title = "{Relationship of EUV Irradiance Coronal Dimming Slope and Depth to Coronal Mass Ejection Speed and Mass}",
      journal = {\apj},
         year = 2016,
        month = oct,
       volume = {830},
       number = {1},
          eid = {20},
        pages = {20},
          doi = {10.3847/0004-637X/830/1/20},
archivePrefix = {arXiv},
       eprint = {1607.05284},
 primaryClass = {astro-ph.SR},
       adsurl = {https://ui.adsabs.harvard.edu/abs/2016ApJ...830...20M}
}

@ARTICLE{2024ApJ...970...60X,
       author = {{Xu}, Yu and {Tian}, Hui and {Veronig}, Astrid M. and {Dissauer}, Karin},
        title = "{Sun-as-a-star Observations of Obscuration Dimmings Caused by Filament Eruptions}",
      journal = {\apj},
         year = 2024,
        month = jul,
       volume = {970},
       number = {1},
          eid = {60},
        pages = {60},
          doi = {10.3847/1538-4357/ad500b},
archivePrefix = {arXiv},
       eprint = {2405.13671},
 primaryClass = {astro-ph.SR},
       adsurl = {https://ui.adsabs.harvard.edu/abs/2024ApJ...970...60X}
}

@ARTICLE{2024A&A...691A.195R,
       author = {{Ronca}, G.~M. and {Chikunova}, G. and {Dissauer}, K. and {Podladchikova}, T. and {Veronig}, A.~M.},
        title = "{Recovery of coronal dimmings}",
      journal = {\aap},
         year = 2024,
        month = nov,
       volume = {691},
          eid = {A195},
        pages = {A195},
          doi = {10.1051/0004-6361/202347934},
archivePrefix = {arXiv},
       eprint = {2410.02585},
 primaryClass = {astro-ph.SR},
       adsurl = {https://ui.adsabs.harvard.edu/abs/2024A&A...691A.195R}
}

@ARTICLE{Gilbert_2007,
       author = {{Gilbert}, Holly R. and {Alexander}, David and {Liu}, Rui},
        title = "{Filament Kinking and Its Implications for Eruption and Re-formation}",
      journal = {\solphys},
         year = 2007,
        month = oct,
       volume = {245},
       number = {2},
        pages = {287-309},
          doi = {10.1007/s11207-007-9045-z},
       adsurl = {https://ui.adsabs.harvard.edu/abs/2007SoPh..245..287G}
}

@ARTICLE{Torok_2004,
       author = {{T{\"o}r{\"o}k}, T. and {Kliem}, B. and {Titov}, V.~S.},
        title = "{Ideal kink instability of a magnetic loop equilibrium}",
      journal = {\aap},
         year = 2004,
        month = jan,
       volume = {413},
        pages = {L27-L30},
          doi = {10.1051/0004-6361:20031691},
archivePrefix = {arXiv},
       eprint = {astro-ph/0311198},
 primaryClass = {astro-ph},
       adsurl = {https://ui.adsabs.harvard.edu/abs/2004A&A...413L..27T}
}

@ARTICLE{Torok_2005,
       author = {{T{\"o}r{\"o}k}, T. and {Kliem}, B.},
        title = "{Confined and Ejective Eruptions of Kink-unstable Flux Ropes}",
      journal = {\apjl},
         year = 2005,
        month = sep,
       volume = {630},
       number = {1},
        pages = {L97-L100},
          doi = {10.1086/462412},
archivePrefix = {arXiv},
       eprint = {astro-ph/0507662},
 primaryClass = {astro-ph},
       adsurl = {https://ui.adsabs.harvard.edu/abs/2005ApJ...630L..97T}
}

@ARTICLE{Thalmann_2015,
       author = {{Thalmann}, J.~K. and {Su}, Y. and {Temmer}, M. and {Veronig}, A.~M.},
        title = "{The Confined X-class Flares of Solar Active Region 2192}",
      journal = {\apjl},
         year = 2015,
        month = mar,
       volume = {801},
       number = {2},
          eid = {L23},
        pages = {L23},
          doi = {10.1088/2041-8205/801/2/L23},
archivePrefix = {arXiv},
       eprint = {1502.05157},
 primaryClass = {astro-ph.SR},
       adsurl = {https://ui.adsabs.harvard.edu/abs/2015ApJ...801L..23T}
}

@ARTICLE{Li_2020,
       author = {{Li}, Ting and {Hou}, Yijun and {Yang}, Shuhong and {Zhang}, Jun and {Liu}, Lijuan and {Veronig}, Astrid M.},
        title = "{Magnetic Flux of Active Regions Determining the Eruptive Character of Large Solar Flares}",
      journal = {\apj},
         year = 2020,
        month = sep,
       volume = {900},
       number = {2},
          eid = {128},
        pages = {128},
          doi = {10.3847/1538-4357/aba6ef},
archivePrefix = {arXiv},
       eprint = {2007.08127},
 primaryClass = {astro-ph.SR},
       adsurl = {https://ui.adsabs.harvard.edu/abs/2020ApJ...900..128L}
}

@ARTICLE{2021NatAs...5..697V,
       author = {{Veronig}, Astrid M. and {Odert}, Petra and {Leitzinger}, Martin and {Dissauer}, Karin and {Fleck}, Nikolaus C. and {Hudson}, Hugh S.},
        title = "{Indications of stellar coronal mass ejections through coronal dimmings}",
      journal = {Nature Astronomy},
         year = 2021,
        month = jan,
       volume = {5},
        pages = {697-706},
          doi = {10.1038/s41550-021-01345-9},
archivePrefix = {arXiv},
       eprint = {2110.12029},
 primaryClass = {astro-ph.SR},
       adsurl = {https://ui.adsabs.harvard.edu/abs/2021NatAs...5..697V}
}

@ARTICLE{Gou_2026,
       author = {{Gou}, Tingyu and {Reeves}, Katharine K. and {Young}, Peter R. and {Veronig}, Astrid M. and {Chen}, Xingyao and {Yu}, Sijie and {Chen}, Bin and {Zhuang}, Bin},
        title = "{Multi-Viewpoint Observation of a Failed Prominence Eruption on the Sun}",
      journal = {Nature Astronomy},
         year = 2026,
        month = may,
          doi = {10.1038/s41550-026-02872-z},
archivePrefix = {arXiv},
       eprint = {2604.23084},
 primaryClass = {astro-ph.SR},
       adsurl = {https://ui.adsabs.harvard.edu/abs/2026arXiv260423084G}
}

@ARTICLE{Namekata_2024,
       author = {{Namekata}, Kosuke and {Airapetian}, Vladimir S. and {Petit}, Pascal and {Maehara}, Hiroyuki and {Ikuta}, Kai and {Inoue}, Shun and {Notsu}, Yuta and {Paudel}, Rishi R. and {Arzoumanian}, Zaven and {Avramova-Boncheva}, Antoaneta A. and {Gendreau}, Keith and {Jeffers}, Sandra V. and {Marsden}, Stephen and {Morin}, Julien and {Neiner}, Coralie and {Vidotto}, Aline A. and {Shibata}, Kazunari},
        title = "{Multiwavelength Campaign Observations of a Young Solar-type Star, EK Draconis. I. Discovery of Prominence Eruptions Associated with Superflares}",
      journal = {\apj},
         year = 2024,
        month = jan,
       volume = {961},
       number = {1},
          eid = {23},
        pages = {23},
          doi = {10.3847/1538-4357/ad0b7c},
archivePrefix = {arXiv},
       eprint = {2311.07380},
 primaryClass = {astro-ph.SR},
       adsurl = {https://ui.adsabs.harvard.edu/abs/2024ApJ...961...23N}
}

@ARTICLE{Mrozek_2020,
       author = {{Mrozek}, Tomasz and {Ko{\l}oma{\'n}ski}, Sylwester and {Ste{\'s}licki}, Marek and {Gronkiewicz}, Dominik},
        title = "{Catalog of Solar Failed Eruptions and Other Dynamic Features Registered by SDO/AIA}",
      journal = {\apjs},
         year = 2020,
        month = aug,
       volume = {249},
       number = {2},
          eid = {21},
        pages = {21},
          doi = {10.3847/1538-4365/ab9e00},
       adsurl = {https://ui.adsabs.harvard.edu/abs/2020ApJS..249...21M}
}

@ARTICLE{2008A&A...478..897B,
       author = {{Bewsher}, D. and {Harrison}, R.~A. and {Brown}, D.~S.},
        title = "{The relationship between EUV dimming and coronal mass ejections. I. Statistical study and probability model}",
      journal = {\aap},
         year = 2008,
        month = feb,
       volume = {478},
       number = {3},
        pages = {897-906},
          doi = {10.1051/0004-6361:20078615},
       adsurl = {https://ui.adsabs.harvard.edu/abs/2008A&A...478..897B}
}

@ARTICLE{2011ApJ...739...89M,
       author = {{Muhr}, N. and {Veronig}, A.~M. and {Kienreich}, I.~W. and {Temmer}, M. and {Vr{\v{s}}nak}, B.},
        title = "{Analysis of Characteristic Parameters of Large-scale Coronal Waves Observed by the Solar-Terrestrial Relations Observatory/Extreme Ultraviolet Imager}",
      journal = {\apj},
         year = 2011,
        month = oct,
       volume = {739},
       number = {2},
          eid = {89},
        pages = {89},
          doi = {10.1088/0004-637X/739/2/89},
archivePrefix = {arXiv},
       eprint = {1107.0921},
 primaryClass = {astro-ph.SR},
       adsurl = {https://ui.adsabs.harvard.edu/abs/2011ApJ...739...89M}
}

@ARTICLE{2015ApJ...811..139T,
       author = {{Tian}, Hui and {Young}, Peter R. and {Reeves}, Katharine K. and {Chen}, Bin and {Liu}, Wei and {McKillop}, Sean},
        title = "{Temporal Evolution of Chromospheric Evaporation: Case Studies of the M1.1 Flare on 2014 September 6 and X1.6 Flare on 2014 September 10}",
      journal = {\apj},
         year = 2015,
        month = oct,
       volume = {811},
       number = {2},
          eid = {139},
        pages = {139},
          doi = {10.1088/0004-637X/811/2/139},
archivePrefix = {arXiv},
       eprint = {1505.02736},
 primaryClass = {astro-ph.SR},
       adsurl = {https://ui.adsabs.harvard.edu/abs/2015ApJ...811..139T}
}

@ARTICLE{2020A&A...633A.142Z,
       author = {{Zhang}, Q.~M. and {Zheng}, R.~S.},
        title = "{Remote coronal dimmings related to a circular-ribbon flare}",
      journal = {\aap},
         year = 2020,
        month = jan,
       volume = {633},
          eid = {A142},
        pages = {A142},
          doi = {10.1051/0004-6361/201937126},
archivePrefix = {arXiv},
       eprint = {1912.09618},
 primaryClass = {astro-ph.SR},
       adsurl = {https://ui.adsabs.harvard.edu/abs/2020A&A...633A.142Z}
}

@Article{Jiang_2022,
title = {Study on complex magnetic structure of solar eruptions},
journal = {Reviews of Geophysics and Planetary Physics (in Chinese)},
volume = {53},
number = {5},
pages = {497-516},
year = {2022},
issn = {2097-1893},
doi = {10.19975/j.dqyxx.2022-022},
url = {https://www.sjdz.org.cn/cn/article/doi/10.19975/j.dqyxx.2022-022},
author = {{Jiang}, Chaowei}
}

@ARTICLE{2025SCPMA..6879611Z,
       author = {{Zheng}, Ruisheng and {Liu}, Yihan and {Zhong}, Ze and {Zhang}, Liang and {Wang}, Xiaoqian and {Li}, Jun and {Wu}, Zhao and {Ying}, Beili and {Feng}, Li and {Su}, Yang and {Gan}, Weiqun and {Hou}, Zhenyong and {Song}, Qiao and {Xue}, Zhike and {Wang}, Jingsong and {Tian}, Hui and {Chen}, Yao},
        title = "{The possible ``double-bang firecracker'' during a solar prominence eruption}",
      journal = {Science China Physics, Mechanics, and Astronomy},
         year = 2025,
        month = jul,
       volume = {68},
       number = {7},
          eid = {279611},
        pages = {279611},
          doi = {10.1007/s11433-024-2648-y},
       adsurl = {https://ui.adsabs.harvard.edu/abs/2025SCPMA..6879611Z}
}

@Article{S22593-lijianping-F,
title = {Diagnostics of temperature and blue-shifted velocity using soft X-ray spectra from the Macau Science Satellite-1},
journal = {Earth and Planetary Physics},
volume = {9},
number = {3},
pages = {740-751},
year = {2025},
issn = {2096-3955},
doi = {10.26464/epp2025036},	
url = {https://www.eppcgs.org/en/article/doi/10.26464/epp2025036},
author = {JianPing Li and Xu Yang and Dong Li and JinHua Shen and Lei Yang and Ya Wang and LianSheng Li and YongQiang Shi and HaiSheng Ji}
}

@ARTICLE{2012SoPh..275...17L,
       author = {{Lemen}, James R. and {Title}, Alan M. and {Akin}, David J. and {Boerner}, Paul F. and {Chou}, Catherine and {Drake}, Jerry F. and {Duncan}, Dexter W. and {Edwards}, Christopher G. and {Friedlaender}, Frank M. and {Heyman}, Gary F. and {Hurlburt}, Neal E. and {Katz}, Noah L. and {Kushner}, Gary D. and {Levay}, Michael and {Lindgren}, Russell W. and {Mathur}, Dnyanesh P. and {McFeaters}, Edward L. and {Mitchell}, Sarah and {Rehse}, Roger A. and {Schrijver}, Carolus J. and {Springer}, Larry A. and {Stern}, Robert A. and {Tarbell}, Theodore D. and {Wuelser}, Jean-Pierre and {Wolfson}, C. Jacob and {Yanari}, Carl and {Bookbinder}, Jay A. and {Cheimets}, Peter N. and {Caldwell}, David and {Deluca}, Edward E. and {Gates}, Richard and {Golub}, Leon and {Park}, Sang and {Podgorski}, William A. and {Bush}, Rock I. and {Scherrer}, Philip H. and {Gummin}, Mark A. and {Smith}, Peter and {Auker}, Gary and {Jerram}, Paul and {Pool}, Peter and {Soufli}, Regina and {Windt}, David L. and {Beardsley}, Sarah and {Clapp}, Matthew and {Lang}, James and {Waltham}, Nicholas},
        title = "{The Atmospheric Imaging Assembly (AIA) on the Solar Dynamics Observatory (SDO)}",
      journal = {\solphys},
         year = 2012,
        month = jan,
       volume = {275},
       number = {1-2},
        pages = {17-40},
          doi = {10.1007/s11207-011-9776-8},
       adsurl = {https://ui.adsabs.harvard.edu/abs/2012SoPh..275...17L}
}

@ARTICLE{2012SoPh..275....3P,
       author = {{Pesnell}, W. Dean and {Thompson}, B.~J. and {Chamberlin}, P.~C.},
        title = "{The Solar Dynamics Observatory (SDO)}",
      journal = {\solphys},
         year = 2012,
        month = jan,
       volume = {275},
       number = {1-2},
        pages = {3-15},
          doi = {10.1007/s11207-011-9841-3},
       adsurl = {https://ui.adsabs.harvard.edu/abs/2012SoPh..275....3P}
}

@ARTICLE{2008SSRv..136....5K,
       author = {{Kaiser}, M.~L. and {Kucera}, T.~A. and {Davila}, J.~M. and {St. Cyr}, O.~C. and {Guhathakurta}, M. and {Christian}, E.},
        title = "{The STEREO Mission: An Introduction}",
      journal = {\ssr},
         year = 2008,
        month = apr,
       volume = {136},
       number = {1-4},
        pages = {5-16},
          doi = {10.1007/s11214-007-9277-0},
       adsurl = {https://ui.adsabs.harvard.edu/abs/2008SSRv..136....5K}
}

@ARTICLE{2008SSRv..136...67H,
       author = {{Howard}, R.~A. and {Moses}, J.~D. and {Vourlidas}, A. and {Newmark}, J.~S. and {Socker}, D.~G. and {Plunkett}, S.~P. and {Korendyke}, C.~M. and {Cook}, J.~W. and {Hurley}, A. and {Davila}, J.~M. and {Thompson}, W.~T. and {St Cyr}, O.~C. and {Mentzell}, E. and {Mehalick}, K. and {Lemen}, J.~R. and {Wuelser}, J.~P. and {Duncan}, D.~W. and {Tarbell}, T.~D. and {Wolfson}, C.~J. and {Moore}, A. and {Harrison}, R.~A. and {Waltham}, N.~R. and {Lang}, J. and {Davis}, C.~J. and {Eyles}, C.~J. and {Mapson-Menard}, H. and {Simnett}, G.~M. and {Halain}, J.~P. and {Defise}, J.~M. and {Mazy}, E. and {Rochus}, P. and {Mercier}, R. and {Ravet}, M.~F. and {Delmotte}, F. and {Auchere}, F. and {Delaboudiniere}, J.~P. and {Bothmer}, V. and {Deutsch}, W. and {Wang}, D. and {Rich}, N. and {Cooper}, S. and {Stephens}, V. and {Maahs}, G. and {Baugh}, R. and {McMullin}, D. and {Carter}, T.},
        title = "{Sun Earth Connection Coronal and Heliospheric Investigation (SECCHI)}",
      journal = {\ssr},
         year = 2008,
        month = apr,
       volume = {136},
       number = {1-4},
        pages = {67-115},
          doi = {10.1007/s11214-008-9341-4},
       adsurl = {https://ui.adsabs.harvard.edu/abs/2008SSRv..136...67H}
}

@ARTICLE{2012SoPh..275..115W,
       author = {{Woods}, T.~N. and {Eparvier}, F.~G. and {Hock}, R. and {Jones}, A.~R. and {Woodraska}, D. and {Judge}, D. and {Didkovsky}, L. and {Lean}, J. and {Mariska}, J. and {Warren}, H. and {McMullin}, D. and {Chamberlin}, P. and {Berthiaume}, G. and {Bailey}, S. and {Fuller-Rowell}, T. and {Sojka}, J. and {Tobiska}, W.~K. and {Viereck}, R.},
        title = "{Extreme Ultraviolet Variability Experiment (EVE) on the Solar Dynamics Observatory (SDO): Overview of Science Objectives, Instrument Design, Data Products, and Model Developments}",
      journal = {\solphys},
         year = 2012,
        month = jan,
       volume = {275},
       number = {1-2},
        pages = {115-143},
          doi = {10.1007/s11207-009-9487-6},
       adsurl = {https://ui.adsabs.harvard.edu/abs/2012SoPh..275..115W}
}

@ARTICLE{2019ApJ...881..151L,
       author = {{Li}, Ting and {Liu}, Lijuan and {Hou}, Yijun and {Zhang}, Jun},
        title = "{Two Types of Confined Solar Flares}",
      journal = {\apj},
         year = 2019,
        month = aug,
       volume = {881},
       number = {2},
          eid = {151},
        pages = {151},
          doi = {10.3847/1538-4357/ab3121},
archivePrefix = {arXiv},
       eprint = {1907.04510},
 primaryClass = {astro-ph.SR},
       adsurl = {https://ui.adsabs.harvard.edu/abs/2019ApJ...881..151L}
}

@MISC{2025kiss.rept.....L,
       author = {{Loyd}, R.~O. Parke and {Shkolnik}, Evgenya L. and {Lazio}, Joseph and {Hallinan}, Gregg W. and {Alvarado-G{\'o}mez}, Juli{\'a}n and {Amaral}, Laura and {Davis}, Ivey and {Farrish}, Alison and {Green}, James and {Brain}, Dave and {Chen}, Bin and {Cohen}, Christina and {Curry}, Shannon and {Dissauer}, Karin and {Egan}, Arika and {Gopalswamy}, Nat and {Gronoff}, Guillaume and {Habbal}, Shadia and {Hu}, Renyu and {Jin}, Meng and {Mason}, James Paul and {Murray-Clay}, Ruth and {Namekata}, Kosuke and {Osten}, Rachel and {Segura}, Ant{\'\i}gona and {Veronig}, Astrid and {Vidotto}, Aline and {Wilson}, Maurice and {Xu}, Yu},
        title = "{The Exospace Weather Frontier}",
 howpublished = {Report prepared for the W. M. Keck Institute for Space Studies (KISS), California Institute of Technology, by R.O.P. Loyd et al, 2025.},
         year = 2025,
        month = oct,
          doi = {10.26206/gmhk5-amp17},
archivePrefix = {arXiv},
       eprint = {2511.02871},
 primaryClass = {astro-ph.IM},
       adsurl = {https://ui.adsabs.harvard.edu/abs/2025kiss.rept.....L}
}

@ARTICLE{2024ScChE..67.1592L,
       author = {{Li}, Dong and {Hou}, ZhenYong and {Bai}, XianYong and {Li}, Chuan and {Fang}, Matthew and {Zhao}, HaiSheng and {Wang}, JinCheng and {Ning}, ZongJun},
        title = "{Simultaneous detection of flare-associated kink oscillations and extreme-ultraviolet waves}",
      journal = {Science in China E: Technological Sciences},
         year = 2024,
        month = may,
       volume = {67},
       number = {5},
        pages = {1592-1601},
          doi = {10.1007/s11431-023-2534-8},
archivePrefix = {arXiv},
       eprint = {2311.08767},
 primaryClass = {astro-ph.SR},
       adsurl = {https://ui.adsabs.harvard.edu/abs/2024ScChE..67.1592L}
}

@ARTICLE{2018ApJ...856L..17S,
       author = {{Su}, Yang and {Veronig}, Astrid M. and {Hannah}, Iain G. and {Cheung}, Mark C.~M. and {Dennis}, Brian R. and {Holman}, Gordon D. and {Gan}, Weiqun and {Li}, Youping},
        title = "{Determination of Differential Emission Measure from Solar Extreme Ultraviolet Images}",
      journal = {\apjl},
         year = 2018,
        month = mar,
       volume = {856},
       number = {1},
          eid = {L17},
        pages = {L17},
          doi = {10.3847/2041-8213/aab436},
       adsurl = {https://ui.adsabs.harvard.edu/abs/2018ApJ...856L..17S}
}

@ARTICLE{2015ApJ...807..143C,
       author = {{Cheung}, Mark C.~M. and {Boerner}, P. and {Schrijver}, C.~J. and {Testa}, P. and {Chen}, F. and {Peter}, H. and {Malanushenko}, A.},
        title = "{Thermal Diagnostics with the Atmospheric Imaging Assembly on board the Solar Dynamics Observatory: A Validated Method for Differential Emission Measure Inversions}",
      journal = {\apj},
         year = 2015,
        month = jul,
       volume = {807},
       number = {2},
          eid = {143},
        pages = {143},
          doi = {10.1088/0004-637X/807/2/143},
archivePrefix = {arXiv},
       eprint = {1504.03258},
 primaryClass = {astro-ph.SR},
       adsurl = {https://ui.adsabs.harvard.edu/abs/2015ApJ...807..143C}
}

@ARTICLE{2011ApJ...739...59W,
       author = {{Woods}, Thomas N. and {Hock}, Rachel and {Eparvier}, Frank and {Jones}, Andrew R. and {Chamberlin}, Phillip C. and {Klimchuk}, James A. and {Didkovsky}, Leonid and {Judge}, Darrell and {Mariska}, John and {Warren}, Harry and {Schrijver}, Carolus J. and {Webb}, David F. and {Bailey}, Scott and {Tobiska}, W. Kent},
        title = "{New Solar Extreme-ultraviolet Irradiance Observations during Flares}",
      journal = {\apj},
         year = 2011,
        month = oct,
       volume = {739},
       number = {2},
          eid = {59},
        pages = {59},
          doi = {10.1088/0004-637X/739/2/59},
       adsurl = {https://ui.adsabs.harvard.edu/abs/2011ApJ...739...59W}
}

@ARTICLE{2018ApJ...863..169D,
       author = {{Dissauer}, K. and {Veronig}, A.~M. and {Temmer}, M. and {Podladchikova}, T. and {Vanninathan}, K.},
        title = "{Statistics of Coronal Dimmings Associated with Coronal Mass Ejections. I. Characteristic Dimming Properties and Flare Association}",
      journal = {\apj},
         year = 2018,
        month = aug,
       volume = {863},
       number = {2},
          eid = {169},
        pages = {169},
          doi = {10.3847/1538-4357/aad3c6},
archivePrefix = {arXiv},
       eprint = {1807.05056},
 primaryClass = {astro-ph.SR},
       adsurl = {https://ui.adsabs.harvard.edu/abs/2018ApJ...863..169D}
}

@ARTICLE{Moore_2001,
       author = {{Moore}, Ronald L. and {Sterling}, Alphonse C. and {Hudson}, Hugh S. and {Lemen}, James R.},
        title = "{Onset of the Magnetic Explosion in Solar Flares and Coronal Mass Ejections}",
      journal = {\apj},
         year = 2001,
        month = may,
       volume = {552},
       number = {2},
        pages = {833-848},
          doi = {10.1086/320559},
       adsurl = {https://ui.adsabs.harvard.edu/abs/2001ApJ...552..833M}
}

@ARTICLE{Ji_2003,
       author = {{Ji}, Haisheng and {Wang}, Haimin and {Schmahl}, Edward J. and {Moon}, Y.-J. and {Jiang}, Yunchun},
        title = "{Observations of the Failed Eruption of a Filament}",
      journal = {\apjl},
         year = 2003,
        month = oct,
       volume = {595},
       number = {2},
        pages = {L135-L138},
          doi = {10.1086/378178},
       adsurl = {https://ui.adsabs.harvard.edu/abs/2003ApJ...595L.135J}
}

@ARTICLE{1997ApJ...491L..55S,
       author = {{Sterling}, Alphonse C. and {Hudson}, Hugh S.},
        title = "{Yohkoh SXT Observations of X-Ray ``Dimming'' Associated with a Halo Coronal Mass Ejection}",
      journal = {\apjl},
         year = 1997,
        month = dec,
       volume = {491},
       number = {1},
        pages = {L55-L58},
          doi = {10.1086/311043},
       adsurl = {https://ui.adsabs.harvard.edu/abs/1997ApJ...491L..55S}
}

@ARTICLE{2018ApJ...857...62V,
       author = {{Vanninathan}, Kamalam and {Veronig}, Astrid M. and {Dissauer}, Karin and {Temmer}, Manuela},
        title = "{Plasma Diagnostics of Coronal Dimming Events}",
      journal = {\apj},
         year = 2018,
        month = apr,
       volume = {857},
       number = {1},
          eid = {62},
        pages = {62},
          doi = {10.3847/1538-4357/aab09a},
archivePrefix = {arXiv},
       eprint = {1802.06152},
 primaryClass = {astro-ph.SR},
       adsurl = {https://ui.adsabs.harvard.edu/abs/2018ApJ...857...62V}
}

@ARTICLE{2016SoPh..291.1761H,
       author = {{Harra}, Louise K. and {Schrijver}, Carolus J. and {Janvier}, Miho and {Toriumi}, Shin and {Hudson}, Hugh and {Matthews}, Sarah and {Woods}, Magnus M. and {Hara}, Hirohisa and {Guedel}, Manuel and {Kowalski}, Adam and {Osten}, Rachel and {Kusano}, Kanya and {Lueftinger}, Theresa},
        title = "{The Characteristics of Solar X-Class Flares and CMEs: A Paradigm for Stellar Superflares and Eruptions?}",
      journal = {\solphys},
         year = 2016,
        month = aug,
       volume = {291},
       number = {6},
        pages = {1761-1782},
          doi = {10.1007/s11207-016-0923-0},
       adsurl = {https://ui.adsabs.harvard.edu/abs/2016SoPh..291.1761H}
}

@ARTICLE{2022ApJ...933...92C,
       author = {{Chen}, Hechao and {Tian}, Hui and {Li}, Hao and {Wang}, Jianguo and {Lu}, Hongpeng and {Xu}, Yu and {Hou}, Zhenyong and {Wu}, Yuchuan},
        title = "{Detection of Flare-induced Plasma Flows in the Corona of EV Lac with X-Ray Spectroscopy}",
      journal = {\apj},
         year = 2022,
        month = jul,
       volume = {933},
       number = {1},
          eid = {92},
        pages = {92},
          doi = {10.3847/1538-4357/ac739b},
archivePrefix = {arXiv},
       eprint = {2205.14293},
 primaryClass = {astro-ph.SR},
       adsurl = {https://ui.adsabs.harvard.edu/abs/2022ApJ...933...92C}
}

@ARTICLE{1970A&A.....6..468L,
       author = {{Landini}, M. and {Monsignori Fossi}, B.~C.},
        title = "{Solar Radiation from 1 to 100 A}",
      journal = {\aap},
         year = 1970,
        month = jul,
       volume = {6},
        pages = {468},
       adsurl = {https://ui.adsabs.harvard.edu/abs/1970A&A.....6..468L}
}

@ARTICLE{1966ApJ...144..244T,
       author = {{Tucker}, W.~H. and {Gould}, R.~J.},
        title = "{Radiation from a Low-Density Plasma at {}10$^{6}$ {\textdegree} - {}10$^{8}$ {\textdegree} K}",
      journal = {\apj},
         year = 1966,
        month = apr,
       volume = {144},
        pages = {244},
          doi = {10.1086/148601},
       adsurl = {https://ui.adsabs.harvard.edu/abs/1966ApJ...144..244T}
}

@ARTICLE{1971ApJ...168..283T,
       author = {{Tucker}, Wallace H. and {Koren}, Marvin},
        title = "{Radiation from a High-Temperature Low-Density Plasma: the X-Ray Spectrum of the Solar Corona}",
      journal = {\apj},
         year = 1971,
        month = sep,
       volume = {168},
        pages = {283},
          doi = {10.1086/151083},
       adsurl = {https://ui.adsabs.harvard.edu/abs/1971ApJ...168..283T}
}

@ARTICLE{2021ApJ...911...33C,
       author = {{Chen}, Hechao and {Yang}, Jiayan and {Hong}, Junchao and {Li}, Haidong and {Duan}, Yadan},
        title = "{Direct Observation of a Large-scale CME Flux Rope Event Arising from an Unwinding Coronal Jet}",
      journal = {\apj},
         year = 2021,
        month = apr,
       volume = {911},
       number = {1},
          eid = {33},
        pages = {33},
          doi = {10.3847/1538-4357/abe6a8},
archivePrefix = {arXiv},
       eprint = {2102.13336},
 primaryClass = {astro-ph.SR},
       adsurl = {https://ui.adsabs.harvard.edu/abs/2021ApJ...911...33C}
}

@ARTICLE{2022SerAJ.205....1L,
       author = {{Leitzinger}, M. and {Odert}, P.},
        title = "{Stellar Coronal Mass Ejections}",
      journal = {Serbian Astronomical Journal},
         year = 2022,
        month = dec,
       volume = {205},
        pages = {1-22},
          doi = {10.2298/SAJ2205001L},
archivePrefix = {arXiv},
       eprint = {2212.09079},
 primaryClass = {astro-ph.SR},
       adsurl = {https://ui.adsabs.harvard.edu/abs/2022SerAJ.205....1L}
}

@ARTICLE{2022ApJ...931...76X,
       author = {{Xu}, Yu and {Tian}, Hui and {Hou}, Zhenyong and {Yang}, Zihao and {Gao}, Yuhang and {Bai}, Xianyong},
        title = "{Sun-as-a-star Spectroscopic Observations of the Line-of-sight Velocity of a Solar Eruption on 2021 October 28}",
      journal = {\apj},
         year = 2022,
        month = jun,
       volume = {931},
       number = {2},
          eid = {76},
        pages = {76},
          doi = {10.3847/1538-4357/ac69d5},
archivePrefix = {arXiv},
       eprint = {2204.11722},
 primaryClass = {astro-ph.SR},
       adsurl = {https://ui.adsabs.harvard.edu/abs/2022ApJ...931...76X}
}

@ARTICLE{2010ApJ...720L..88R,
       author = {{Robbrecht}, Eva and {Wang}, Yi-Ming},
        title = "{The Temperature-dependent Nature of Coronal Dimmings}",
      journal = {\apjl},
         year = 2010,
        month = sep,
       volume = {720},
       number = {1},
        pages = {L88-L92},
          doi = {10.1088/2041-8205/720/1/L88},
archivePrefix = {arXiv},
       eprint = {1007.5191},
 primaryClass = {astro-ph.SR},
       adsurl = {https://ui.adsabs.harvard.edu/abs/2010ApJ...720L..88R}
}

@ARTICLE{2014ApJ...789...61M,
       author = {{Mason}, James Paul and {Woods}, T.~N. and {Caspi}, A. and {Thompson}, B.~J. and {Hock}, R.~A.},
        title = "{Mechanisms and Observations of Coronal Dimming for the 2010 August 7 Event}",
      journal = {\apj},
         year = 2014,
        month = jul,
       volume = {789},
       number = {1},
          eid = {61},
        pages = {61},
          doi = {10.1088/0004-637X/789/1/61},
archivePrefix = {arXiv},
       eprint = {1404.1364},
 primaryClass = {astro-ph.SR},
       adsurl = {https://ui.adsabs.harvard.edu/abs/2014ApJ...789...61M}
}

@ARTICLE{2019ApJ...884L..13A,
       author = {{Alvarado-G{\'o}mez}, Juli{\'a}n D. and {Drake}, Jeremy J. and {Moschou}, Sofia P. and {Garraffo}, Cecilia and {Cohen}, Ofer and {NASA LWS Focus Science Team: Solar-Stellar Connection} and {Yadav}, Rakesh K. and {Fraschetti}, Federico},
        title = "{Coronal Response to Magnetically Suppressed CME Events in M-dwarf Stars}",
      journal = {\apjl},
         year = 2019,
        month = oct,
       volume = {884},
       number = {1},
          eid = {L13},
        pages = {L13},
          doi = {10.3847/2041-8213/ab44d0},
archivePrefix = {arXiv},
       eprint = {1909.04092},
 primaryClass = {astro-ph.SR},
       adsurl = {https://ui.adsabs.harvard.edu/abs/2019ApJ...884L..13A}
}
\bibliographystyle{aasjournalv7}

\end{document}